\documentclass[a4paper]{jpconf}
\usepackage{graphicx}
\usepackage{indentfirst}
\usepackage{amsmath}
\begin{document}
\title{Search for $t\bar{t}$ resonances in the semileptonic final state using pp collisions at $\sqrt s$~=~8~\,TeV recorded with the CMS detector}

\author{Anne-Laure Pequegnot}

\address{Institut de Physique Nucl\'eaire de Lyon, France}

\ead{al.pequegnot@ipnl.in2p3.fr}

\begin{abstract}

Many extensions to the Standard Model predict gauge interactions with enhanced couplings to third generation quarks, especially the extremely heavy top quark\cite{theoretical}. These phenomena can lead to resonances in the production of $t\bar{t}$ pairs. In this note, a model-independent search for the production of heavy resonances decaying into top-antitop quark pairs in the semileptonic final state using pp collisions at $\sqrt s$ = 8 TeV recorded with the CMS detector\cite{PRL} is presented. 

The analysis focuses on the semileptonic decay channels into electrons or muons and covers the range of 0.5-3 TeV in the invariant mass of the $t\bar{t}$ pairs where the events present specific and different topologies. Therefore two dedicated analyses have been performed to achieve the best sensitivity on the whole invariant mass range: the threshold analysis (0.5-2 TeV) which is optimized for $t\bar{t}$ production at the kinematic production threshold, using standard top quark reconstruction techniques, and the boosted analysis (1-3 TeV) optimized for $t\bar{t}$ produced with high Lorentz boosts, using dedicated techniques for boosted top quarks reconstruction. 

No evidence of signal has been found. Thus the following limits at 95\% C.L. on the production of non-SM particles in specific models are set: topcolor Z$^{\prime}$ bosons with a width of 1.2 (10) \% of its mass are excluded for masses below 2.1 TeV (up to 2.7 TeV). In addition, Kaluza-Klein excitations of a gluon with masses below 2.5 TeV in the Randall-Sundrum model are excluded.
\end{abstract}

\section{Event selection}

Data samples corresponding to an integrated luminosity of 19.6 $fb^{-1}$ of pp collisions at $\sqrt s$~=~8~\,TeV collected by the CMS experiment in 2012 are analyzed. The CMS detector, a general-purpose apparatus operating at the CERN LHC, is described in detail elsewhere \cite{CMS}.

\medskip

Concerning the threshold analysis (inclusive mass range of 0.5-2 TeV in $m_{t\bar{t}}$), standard reconstruction techniques for top quarks produced with a small boost (all decay products are resolved) are used. The data analysed are recorded with triggers requiring one isolated lepton and 3 central jets. The following selection is applied: exactly one isolated lepton, at least 4 jets with $p_{T} > 70/50/30/30 \,GeV$, $E_{T}^{miss} >$ 20 GeV to reduce QCD background and at least one b-tagged jet.

Events are classified into 4 categories according to the flavour of the lepton and the number of b-tagged jets (exactly one or two and more).

\medskip

For the boosted analysis (inclusive mass range of 1-3 TeV in $m_{t\bar{t}}$), techniques dedicated to boosted top quarks reconstruction are used. The data analysed are recorded with triggers requiring one lepton and 2 central jets. For the selection, exactly one lepton without any isolation specification and at least 2 jets with $p_{T} >$ 150 GeV for the leading jet and $p_{T} >$ 50 GeV for the subsequent jets are required. $E_{t}^{miss}$ has to be greater than 50 GeV to reduce QCD background, and the scalar quantity $T_{T}^{miss} = E_{T}^{miss} + p_{T}^{lepton}$ greater than  150 GeV. To reduce QCD background, especially in electron channel, several topological requirements are applied to ensure the missing transverse momentum does not point along the transverse direction of the electron or of the leading jet. Events are classified into 4 categories according to the flavour of the lepton and the number of b-tagged jets (exactly zero or one and more).

\section{Event reconstruction}

\subsection{Objects assignement}
A first step consists in assigning the final-state objects in each event to either the leptonic or hadronic side of the $t\bar{t}$ decay. the charged lepton and $E_{T}^{miss}$ are assigned to the leptonic side of the event, where $E_{T}^{miss}$ is assigned to the transverse component of the neutrino’s momentum. After that, a $\chi ^{2}$ sorting algorithm is used to assign the jets.

In the threshold analysis, the $\chi ^{2}$ is defined as follows:
\begin{equation}
\chi^{2} = \chi^{2}_{m_t^{Lep}} + \chi^{2}_{m_t^{Had}} + \chi^{2}_{m_w^{Had}} + \chi^{2}_{p_{T}^{t\bar{t}}} \;\;\text{with}\;\; \chi^{2}_{x} = (x_{meas}-x_{MC})^{2}/\sigma _{MC}^{2}
\end{equation}
$m_t^{Lep}$ is the mass of the leptonic top quark ; $m_t^{Had}$, the mass of the hadronic top quark ; $m_w^{Had}$, the mass of the hadronic W boson and $p_{T}^{t\bar{t}}$, the transverse momentum of the $t\bar{t}$ system. The choice of the best jets combination corresponds to the one which gives the smallest $\chi ^{2}$ value.

\medskip
In the boosted analysis, the $\chi ^{2}$ is reduced at its two first terms. All mass hypotheses that have exactly one jet assigned to the leptonic side, and at least one jet assigned to the hadronic side are then considered. The combination with the smallest value of  $\chi ^{2}$  is chosen for each event. This  $\chi ^{2}$ has to be smaller than 10 which rejects most of the W+jet background and maximizes the sensitivity on the expected limits.

\subsection{Background estimation and parametrization}

\begin{figure}[h]
\begin{center}
\begin{minipage}{17.7pc}
\includegraphics[width=16pc]{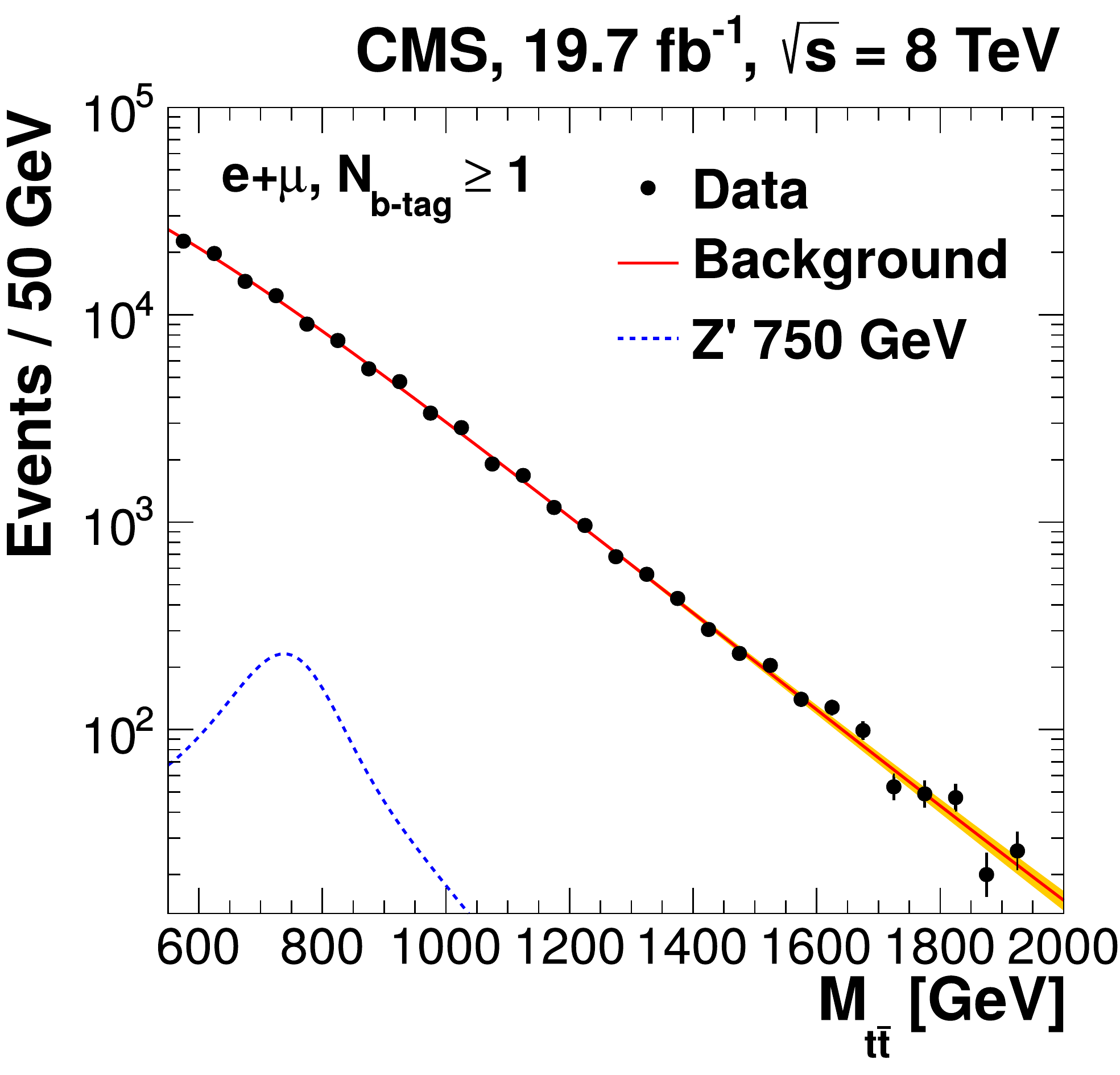}
\caption{\label{Threshold analysis}Likelihood fit projection on data, including a signal with a mass hypothesis of 750 GeV, done simultaneously on the 4 different categories for the threshold analysis. The expected shape of the Z$^{\prime}$ signal is overlaid, normalized to 1.0 pb.}
\end{minipage}\hspace{2pc}%
\begin{minipage}{17.7pc}
\includegraphics[width=16pc]{./semileptonicBoostedMu1b_2.pdf}
 \caption{\label{Boosted analysis}Comparison between data and SM prediction for reconstructed $m_{t\bar{t}}$ distributions for the boosted semileptonic analysis with muon and  $\geq $ 1 b-tagged jets, including a signal with a mass hypothesis of 2 TeV. A cross section of 1.0 pb is used for the normalization of the Z$^{\prime}$  samples.}
\end{minipage} 
\end{center}
\end{figure}

Two particular models for the signal are considered: a generic spin 1 Z$^{\prime}$ boson with a width of 1.2 \% (10 \%) of the resonance mass \cite{topcolorZprime};
and the Kaluza-Klein partner of the SM gluon \cite{kkgluons}.

In the threshold analysis, the signal contribution is extracted from a maximum likelihood fit to the data, using simultaneously the four categories described previously (see figure \ref{Threshold analysis}). The signal Probability Density Function in the fit is parametrized from the simulation, using a superposition of Gaussian kernels to Tmodel the distribution. the background is estimated on data using the following functional form:

\begin{equation}
\frac{d\sigma}{dm_{t\bar{t}}} = \frac{\left(1-\frac{m}{\sqrt{s}}\right)^{c_{1}}}{\left(\frac{m}{\sqrt{s}}\right)^{c_{2}+c_{3}\ln \frac{m}{\sqrt{s}}}}
\end{equation}

The $c_{1}, c_{2}, c_{3}$ parameters and the New Physics signal cross-section are floatting and are extracted directly from the fit on data.

\medskip
In the boosted analysis, the background is estimated from simulations (see figure \ref{Boosted analysis}). 
To extract the number of signal events,  a binned likelihood of the $m_{t\bar{t}}$ distributions in the four channels is used. The number of events in bin i is assumed to follow a Poisson distribution with mean $\lambda _{i}$, given by the sum over all considered background processes and the signal. The signal is scaled with a signal strength modifier $\mu $, which corresponds to the signal cross section in pb:
\begin{equation}
\lambda _{i} = \mu S_{i} + \sum_{k} B_{k,i}
\end{equation}
where k runs over all considered background processes, $B_{k}$ is the background template for background k, and S is the signal template, scaled according to luminosity and a signal cross section of 1 pb.

\section{Systematic uncertainties}

 The treatment of the systematic uncertainties varies between the threshold and boosted analysis owing to the differences in the techniques. The table \ref{syst} summarizes the systematic uncertainties considered, specifying if they affect the signal, the background or both.

\begin{center}
\begin{table}[h]
\footnotesize 
\centering
\begin{tabular}[width=\textwidth]{lcccc}
\hline
 & \multicolumn{2}{c}{\begin{bf}Threshold analysis\end{bf}} & \multicolumn{2}{c}{\begin{bf}Boosted analysis\end{bf}}\\
\begin{bf}Systematic uncertainty\end{bf} & signal & background & signal & background\\
\hline
Event pileup & x &  & x & x \\
Luminosity & x &  & x & x \\
Lepton ID and trigger & x &  & x & x \\
JES and JER & x &  & x & x \\
Signal fit pdf & x &  &  &  \\
Background fit pdf &  & x &  &  \\
Background cross section &  &  &  & x \\
Parton distribution functions & negl. &  & x & x \\
Background modelling &  &  &  & x \\
\hline
\end{tabular} 
\caption[Summary of systematic uncertainties considered.]{Summary of systematic uncertainties considered. A cross indicates when a certain systematic uncertainty was applied, "negl." indicates that the systematic uncertainty was found to be negligible.}
\label{syst}
\end{table}
\end{center}

\section{Results}

The statistical treatment doesn't indicate any excess of events above the expected yield of the SM processes, thus limits are set. A bayesian statistical method is used to extract the 95\% C.L. upper limits. The systematic uncertainties taken into account as nuisance parameters, integrated with a log normal prior.
The transition between the two analyses is based on the expected limits sensitivity. An analysis for the all-hadronic channel has also been performed (only in the boosted regime), but is not described here (see \cite{allHadronic} for more details). This analysis is combined with the semileptonic boosted one.

\begin{figure}[h]
\begin{center}
\includegraphics[width=16pc]{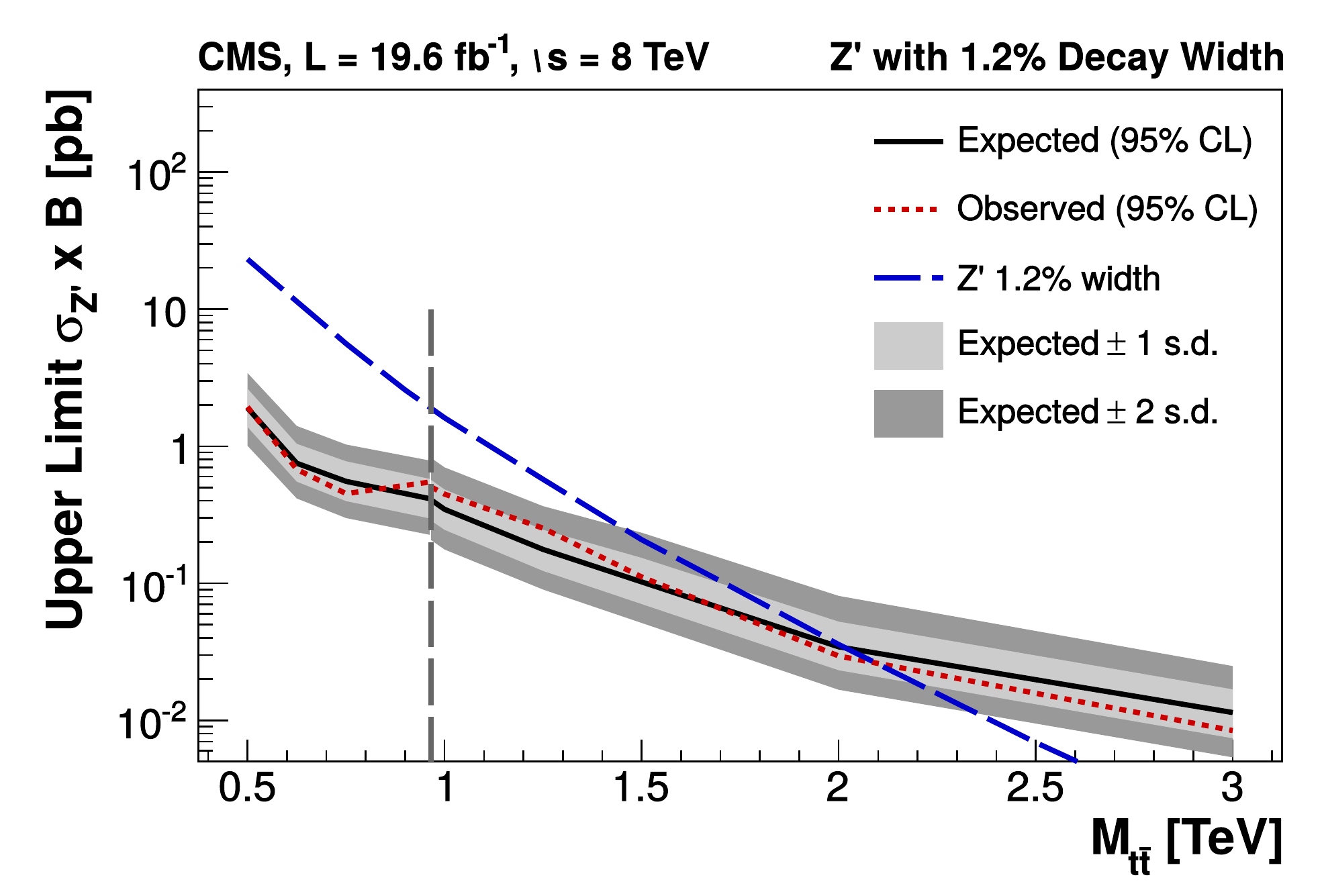}
\caption{\label{Result of the combination}Final result of the combination of the 3 analysis: the 95\% CL upper limits on the product of the production cross section $\sigma _{Z^{\prime}}$ and the branching fraction B of hypothesized resonances that decay into $t\bar{t}$ as a function of the invariant mass of the resonance. The Z$^{\prime}$ production with $\Gamma _{Z^{\prime}}/m_{Z^{\prime}}$ = 1.2\% compared to predictions based on Ref. \cite{topcolorZprime}. The $\pm  1$ and $\pm  2$ s.d. exclusions from the expected limits are also shown.}
\end{center}
\end{figure}

All these combinations lead to the following limits on the production of non-SM particles in specific models. Topcolor Z$^{\prime}$ bosons with a width of 1.2 \% (10\%) of its mass are excluded at 95\% CL for masses below 2.1 TeV (2.7 TeV) (see figure \ref{Result of the combination} for the Z$^{\prime}$ with narrow decay width). Kaluza-Klein excitations of a gluon with masses below 2.5 TeV in the Randall-Sundrum model are excluded.

\end{document}